\RequirePackage{fix-cm}
\documentclass[11pt]{article}
\usepackage[margin=1in]{geometry}

\usepackage[T1]{fontenc}
\usepackage[utf8]{inputenc}
\usepackage{amsmath,amssymb,amsfonts,amsthm,mathtools}
\usepackage{array,cite,setspace,float}
\usepackage{authblk}
\usepackage{bm}
\usepackage{dsfont}
\usepackage{booktabs}
\usepackage{enumitem}
\usepackage{graphicx}
\usepackage{subcaption}
\usepackage[numbers,sort&compress,square]{natbib}
\usepackage[colorlinks=true,citecolor=blue,linkcolor=blue,urlcolor=blue,breaklinks=true]{hyperref}
\usepackage{tikz}
\usepackage{newtxtext,newtxmath}
\usepackage{xcolor}

\title{Long-wavelength density response and momentum resolution in strange-metal charge spectroscopy}
\author{Ilmo Sung\thanks{\texttt{ilmo.sung@hq.dhs.gov}}}
\affil{Science and Technology Directorate, \\ U.S.\ Department of Homeland Security, Washington, DC 20032, USA }
\date{}
\begin{document}
\maketitle
\vspace{-2.5em}
\begin{abstract}
Momentum-resolved electron energy-loss spectroscopy (EELS) on strange-metal Bi$_2$Sr$_2$CaCu$_2$O$_{8+x}$ (Bi-2212) observes a broad charge continuum that is nearly momentum independent over an extended finite-momentum regime, whereas optical spectroscopy determines a metallic local conductivity.  We analyze the long-wavelength response function that connects these regimes: the proper homogeneous bulk charge-density polarization.  For any homogeneous $U(1)$-conserving metal whose longitudinal matter conductivity remains finite in the $q\to0$ limit at fixed nonzero frequency, the Ward identity gives $\chi''_{\rho\rho}({\bf q},\omega)=q^2\operatorname{Re}\sigma_L({\bf q},\omega)/\omega$.  We then prove that any nonnegative normalized momentum-resolution kernel with a self-similar $q$-scaled profile, uniformly bounded second moment, and tight second-moment tails preserves this $q^2$ asymptotic behavior, changing only the prefactor.  Consequently, a nearly local finite-$q$ continuum in Bi-2212, if continuously connected to a regular metallic optical limit of the same proper bulk response, is naturally described by a crossover surface $q_\ast(\omega,T)$.  The theorem is a constraint on the proper bulk density response, or on a positive $q$-scaled convolution of it; screened loss functions and surface EELS observables require the corresponding electrodynamic conversion before the constraint is applied.
\end{abstract}


\section{Introduction}

Momentum-resolved charge spectroscopy of strange-metal Bi$_2$Sr$_2$CaCu$_2$O$_{8+x}$ (Bi-2212) has revealed unusual finite-momentum charge dynamics, and the experimental literature contains both plasmonic and continuum observations.  Early transmission electron energy-loss spectroscopy (EELS) studies of cuprates reported dispersive, comparatively well-defined plasmons at long wavelength~\cite{Nucker1991,Grigoryan1999}.  More recent low-energy momentum-resolved EELS (M-EELS) measurements established a broad finite-momentum continuum in Bi-2212 and followed its evolution across the strange-metal phase diagram~\cite{Vig2017,Mitrano2018,Husain2019}.  The interpretation of these results has been debated, with a Comment arguing for a more conventional plasmon and band-theory interpretation and a Reply emphasizing the persistence of the finite-$q$ continuum beyond a low-momentum plasmonic regime~\cite{Fink2021,HusainReply2021}.  A recent review by Abbamonte and Fink summarizes this broader EELS context~\cite{AbbamonteFink2025}.  Recent reflection M-EELS measurements report a small-momentum regime consistent with plasmonic and optical phenomenology, together with an incoherent and nearly momentum-independent continuum at larger momenta~\cite{Chen2024}.  Transmission EELS measurements likewise report a highly damped low-momentum excitation and a continuum only beyond a finite momentum scale~\cite{deVries2026}.

This article takes the finite-$q$ continuum and the long-wavelength optical response as simultaneous empirical inputs.  The key question is therefore not whether a finite-momentum continuum or a plasmonic low-$q$ feature exists in some experimental window.  The key question is how the finite-$q$ charge response crosses over to the density response implied by a metallic optical conductivity.  The same issue arises more broadly in bad or strongly correlated metals, where some measurements report overdamped plasmons whereas others find propagating collective modes; for example, recent transmission EELS work on Sr$_2$RuO$_4$ reports well-defined optical and acoustic plasmons in a bad-metal setting~\cite{Schultz2025}.  The framework below focuses on the response-function connection between these regimes rather than on microscopic fits to each lineshape.  It isolates one universal constraint on the proper bulk charge-density response.

The response functions involved in this question are distinct.  Optical spectroscopy determines the $q\simeq0$ optical conductivity; under the usual local-electrodynamic identification, its regular part is the $q\to0$ longitudinal conductivity entering the Ward identity.  This identification assumes the standard local-electrodynamic limit in which the regular longitudinal and transverse conductivities coincide as $q\to0$ at fixed nonzero frequency.  EELS experiments measure geometry-dependent loss functions in reflection or transmission geometries, and the extraction of a bulk density response requires an electrodynamic model of the geometry~\cite{Chen2024,deVries2025}.  The result derived below addresses the bulk consistency problem that remains after this conversion has been specified: for a single homogeneous bulk longitudinal response that conserves charge and whose longitudinal conductivity connects regularly to optics, charge conservation fixes the leading small-$q$ density response.

The Ward-identity statement that a conserved density response is suppressed at small momentum is standard.  The added result here concerns momentum resolution.  We show that a broad class of nonnegative $q$-scaled momentum-resolution kernels can modify prefactors and finite-window line shapes while preserving the leading $q^2$ power required by charge conservation and a finite optical conductivity.  Thus a positive momentum-resolution model whose width shrinks in proportion to the nominal momentum cannot change the asymptotic power law of the proper conserved density response.

Accordingly, the article provides a theorem-level consistency result for response functions, with Bi-2212 as the motivating application.  Its experimentally useful output is the crossover surface $q_\ast(\omega,T)$ separating the conserved optical regime from any nearly local finite-momentum regime.

By ``asymptotically local'' we mean that, at fixed nonzero frequency, the proper longitudinal charge-density spectral weight tends to a nonzero $q^0$ function as $q\to0$.  For a conserved homogeneous metal with a finite $q\to0$ longitudinal matter conductivity, the density spectral weight instead vanishes as $q^2$.  The theorem below states that positive self-similar momentum averaging whose width shrinks in proportion to the nominal momentum preserves that power law.

For Bi-2212 this gives a conservative interpretation of the finite-$q$ continuum.  The continuum can be a genuine feature of the experimentally accessible strange-metal regime, while the strict long-wavelength proper density response still crosses over to the conserved optical behavior.  The purpose of this article is to state the constraint precisely, prove the momentum-resolution theorem, separate it from fixed-width resolution scenarios, discuss possible physical origins of the crossover scale, and identify the assumptions needed to use the result in charge-spectroscopy analyses.

\section{Response functions, geometry, and order of limits}
\label{sec:conventions}

Because longitudinal charge spectroscopy in a charged metal involves both matter response and long-range electrodynamics, it is important to specify the response function constrained here.  Throughout this article $\chi^R_{\rho\rho}({\bf q},\omega)$ denotes the irreducible, or proper, homogeneous bulk longitudinal charge-density polarization: the response of the charge density $\rho$ to the total scalar electric potential acting at the material before Coulomb self-consistency, surfaces, and geometry have been iterated into a measured loss function.  This is the object related by the continuity equation to the matter longitudinal conductivity $\sigma_L({\bf q},\omega)$ in standard linear-response and electron-liquid conventions~\cite{Kubo1957,FetterWalecka1971,Mahan2000,GiulianiVignale2005}.  Equivalently, $\chi_{\rho\rho}$ is the polarization entering the longitudinal dielectric function before the final Coulomb resummation, not the fully screened response to an external electron.  If one instead uses the number density $n$, with $\rho=-en$, then $\chi''_{\rho\rho}=e^2\chi''_{nn}$ when the scalar source is converted consistently.  The small-$q$ power law is unchanged.

In the usual local optical limit, the regular $q\to0$ part of the longitudinal matter conductivity is the same conductivity inferred from in-plane optical spectroscopy.  That local-electrodynamic identification is the sense in which optical spectroscopy enters the theorem.  Coulomb propagation, sample geometry, and boundary conditions are accounted for in the separate electrodynamic step that maps a measured loss signal to the bulk proper density response or to a positive convolution of it.

The wave vector entering the Ward identity is the wave vector of the conserved density mode.  In the quasi-two-dimensional notation used below, $q$ denotes the in-plane component of the layer-uniform bulk response, or a component for which any perpendicular or interlayer momentum has already been projected out and does not remain finite as $q\to0$.  Measurements that sample a fixed perpendicular momentum, a surface evanescent field, or a geometry-dependent distribution of out-of-plane momenta require their own electrodynamic reduction before they can be compared with the two-dimensional $q$-scaled convolution considered below.

This distinction is essential.  A raw loss function such as $-\operatorname{Im}\epsilon_L^{-1}$ can have a different small-$q$ power from $\chi''_{\rho\rho}$, because Coulomb kernels themselves carry powers of $q$.  For example, in an ideal three-dimensional continuum the Coulomb interaction scales as $V_q\propto q^{-2}$, so a density response proportional to $q^2$ can produce a loss function with a finite long-wavelength limit.  A nearly $q$-independent loss signal is therefore consistent with the Ward identity when the relevant Coulomb and boundary-condition factors are retained.  The theorem constrains the proper density continuum in a homogeneous conserved bulk response.  This separation between density response, dielectric response, and measured EELS intensity is standard in electron-energy-loss analysis~\cite{Egerton2011,Vig2017,Chen2024,deVries2025}.

We also fix the order of limits.  The Ward constraint used here is a statement at fixed $\omega>0$ followed by $q\to0$.  Static response, singular zero-frequency Drude or superfluid weights, and paths such as $\omega\sim Dq^2$ in a diffusive metal involve different limits because hydrodynamic denominators mix $q$ and $\omega$.  The Bi-2212 question considered here is whether a finite-frequency, finite-$q$ continuum connects to the fixed-frequency local optical limit as a $q^0$ proper density response or through a crossover to the conserved $q^2$ regime.

\section{Ward identity for the proper charge-density response}
\label{sec:ward}

Let $\Phi({\bf q},\omega)$ be the total longitudinal scalar electric potential used to define the matter conductivity, and let
\begin{equation}
\delta\rho({\bf q},\omega)
=\chi^R_{\rho\rho}({\bf q},\omega)\Phi({\bf q},\omega).
\end{equation}
With Fourier convention $e^{i({\bf q}\cdot{\bf r}-\omega t)}$, the continuity equation and longitudinal Ohm law are
\begin{equation}
-i\omega\,\delta\rho({\bf q},\omega)+i{\bf q}\cdot{\bf J}({\bf q},\omega)=0,
\end{equation}
\begin{equation}
{\bf J}_L({\bf q},\omega)=\sigma_L({\bf q},\omega){\bf E}_L({\bf q},\omega),
\qquad
{\bf E}_L=-i{\bf q}\Phi.
\end{equation}
Combining these relations gives
\begin{equation}
\chi^R_{\rho\rho}({\bf q},\omega)
=
\frac{q^2}{i\omega}\sigma_L({\bf q},\omega).
\label{eq:ward}
\end{equation}
For $\omega>0$ we define the positive dissipative density spectrum as
\begin{equation}
\chi''_{\rm bulk}({\bf q},\omega)
\equiv
-\operatorname{Im}\chi^R_{\rho\rho}({\bf q},\omega).
\end{equation}
Then Eq.~\eqref{eq:ward} implies the exact fixed-frequency relation
\begin{equation}
\chi''_{\rm bulk}({\bf q},\omega)
=
\frac{q^2}{\omega}\operatorname{Re}\sigma_L({\bf q},\omega).
\label{eq:exact}
\end{equation}
For an anisotropic conductor this becomes
\begin{equation}
\chi''_{\rm bulk}({\bf q},\omega)
=
\frac{q_iq_j}{\omega}\operatorname{Re}\sigma_{ij}({\bf q},\omega),
\label{eq:tensor-exact}
\end{equation}
with repeated spatial indices summed.

If $\operatorname{Re}\sigma_L({\bf q},\omega)$ has a finite limit as ${\bf q}\to0$ at fixed $\omega>0$, Eq.~\eqref{eq:exact} gives
\begin{equation}
\chi''_{\rm bulk}({\bf q},\omega)
=
q^2 S(\omega)+o(q^2),
\qquad
S(\omega)=\frac{\operatorname{Re}\sigma_{\rm opt}(\omega)}{\omega},
\label{eq:bulk-asymptotic}
\end{equation}
where the optical conductivity is understood in the local-electrodynamic sense described above.  More generally, for any compact frequency interval $I\subset(0,\infty)$ on which the conductivity limit is uniform,
\begin{equation}
r({\bf q},\omega)
\equiv
\chi''_{\rm bulk}({\bf q},\omega)-q^2S(\omega)
\end{equation}
obeys
\begin{equation}
\sup_{\omega\in I}\frac{|r({\bf q},\omega)|}{q^2}\to0
\qquad ({\bf q}\to0).
\label{eq:remainder}
\end{equation}
Equation~\eqref{eq:bulk-asymptotic} is the essential asymptotic statement.  It is independent of quasiparticles, Boltzmann transport, or any particular model of the strange metal.  A putative $q^0$ proper density continuum at fixed nonzero $\omega$ would require $\operatorname{Re}\sigma_L({\bf q},\omega)\sim q^{-2}$ as $q\to0$, which is precisely the failure of a finite long-wavelength conductivity.

\subsection{Diffusive example and order of limits}
\label{sec:hydro-example}

A diffusive conductor illustrates the fixed-frequency nature of the statement.  Let $\kappa_\rho>0$ denote the magnitude of the static charge compressibility in this source convention.  With the present sign convention the static response is $\chi^R_{\rho\rho}(q,0)=-\kappa_\rho$.  The standard hydrodynamic response~\cite{Forster1975} can be written as
\begin{equation}
\chi^R_{\rho\rho}(q,\omega)
=
-\frac{\kappa_\rho Dq^2}{Dq^2-i\omega},
\label{eq:diffusive-response}
\end{equation}
so that
\begin{equation}
\chi''_{\rho\rho}(q,\omega)
=
\frac{\kappa_\rho Dq^2\omega}{\omega^2+(Dq^2)^2}.
\label{eq:diffusive-spectrum}
\end{equation}
At fixed nonzero $\omega$, this is proportional to $q^2$, consistent with Eq.~\eqref{eq:exact} and the Einstein relation $\sigma=D\kappa_\rho$.  Along the hydrodynamic scaling path $\omega\sim Dq^2$, however, the response is controlled by the diffusion pole and the apparent power counting changes.  This example illustrates why the fixed-frequency optical-limit Ward statement and finite-$q$ hydrodynamic scaling are separate limits.

\section{A theorem for \texorpdfstring{$q$}{q}-scaled momentum-resolution kernels}
\label{sec:smearing}

Experimental spectra are never measured with infinite momentum resolution.  In addition, surface roughness, disorder, domains, or other inhomogeneous effects can randomize momentum.  The question addressed here is narrower than arbitrary disorder or electrodynamic mode mixing: can a nonnegative convolution model of momentum resolution change the conserved $q^2$ proper density response into an observed $q^0$ density continuum as $q\to0$ if the resolution width itself scales as $q$?

The theorem gives a definite answer for this kernel class.  It is stated for the in-plane response relevant to the layer-uniform component described in Sec.~\ref{sec:conventions}; arbitrary unresolved out-of-plane momentum averaging requires a separate reduction before comparison with the two-dimensional kernel below.  For notational simplicity the theorem is written in the isotropic scalar form of Eq.~\eqref{eq:bulk-asymptotic}; in an anisotropic metal, $S(\omega)|{\bf Q}|^2$ is replaced by $Q_iQ_j\operatorname{Re}\sigma_{ij}(\omega)/\omega$, which changes only the angular prefactor and not the leading $q^2$ conclusion.

\paragraph*{Proposition.}
Let $I\subset(0,\infty)$ be compact.  Suppose that the proper bulk charge-density spectral function obeys
\begin{equation}
\chi''_{\rm bulk}({\bf Q},\omega)
=
|{\bf Q}|^2S(\omega)+r({\bf Q},\omega),
\end{equation}
with the uniform remainder condition in Eq.~\eqref{eq:remainder}.  Let the measured quantity at nominal in-plane momentum $q\hat{\bf q}$ be the positive convolution
\begin{equation}
\chi''_{\rm meas}(q\hat{\bf q},\omega)
=
\int_{\mathbb{R}^2}d^2Q\,K_{q,\hat{\bf q}}({\bf Q})
\chi''_{\rm bulk}({\bf Q},\omega),
\label{eq:measured}
\end{equation}
where
\begin{equation}
K_{q,\hat{\bf q}}({\bf Q})
=
\frac{1}{q^2}k_{\hat{\bf q}}\!\left(
\frac{{\bf Q}-q\hat{\bf q}}{q}
\right)
\label{eq:qscaled-kernel}
\end{equation}
is nonnegative and normalized,
\begin{equation}
k_{\hat{\bf q}}({\bf x})\ge0,
\qquad
\int_{\mathbb{R}^2}d^2x\,k_{\hat{\bf q}}({\bf x})=1.
\end{equation}
Assume a uniformly bounded second moment,
\begin{equation}
\sup_{\hat{\bf q}\in S^1}
\int_{\mathbb{R}^2}d^2x\,|{\bf x}|^2 k_{\hat{\bf q}}({\bf x})<\infty,
\label{eq:second-moment}
\end{equation}
and uniformly tight second-moment tails,
\begin{equation}
\sup_{\hat{\bf q}\in S^1}
\int_{|{\bf x}|>R}d^2x\,|{\bf x}|^2 k_{\hat{\bf q}}({\bf x})\to0
\qquad (R\to\infty).
\label{eq:tight-tails}
\end{equation}
Finally, assume the ultraviolet domination condition
\begin{equation}
\sup_{{\bf Q}\in\mathbb{R}^2,\,\omega\in I}
\frac{|\chi''_{\rm bulk}({\bf Q},\omega)|}{1+|{\bf Q}|^2}
\le A_I<\infty.
\label{eq:growth}
\end{equation}
Then
\begin{equation}
\chi''_{\rm meas}(q\hat{\bf q},\omega)
=
q^2 C(\hat{\bf q})S(\omega)+o(q^2),
\label{eq:theorem}
\end{equation}
uniformly for $\omega\in I$, with
\begin{equation}
C(\hat{\bf q})=
\int_{\mathbb{R}^2}d^2x\,|\hat{\bf q}+{\bf x}|^2 k_{\hat{\bf q}}({\bf x}).
\label{eq:C}
\end{equation}

The ultraviolet condition in Eq.~\eqref{eq:growth} is only used to control the far tails of the kernel after the change of variables ${\bf Q}=q(\hat{\bf q}+{\bf x})$.  In a lattice implementation it may be replaced by boundedness on the relevant Brillouin-zone domain, or by the corresponding periodic/form-factor bound.

For a centered kernel,
\begin{equation}
\int d^2x\,{\bf x}\,k_{\hat{\bf q}}({\bf x})=0,
\label{eq:centered}
\end{equation}
Eq.~\eqref{eq:C} reduces to
\begin{equation}
C(\hat{\bf q})=1+M_2(\hat{\bf q}),
\qquad
M_2(\hat{\bf q})=\int d^2x\,|{\bf x}|^2 k_{\hat{\bf q}}({\bf x}).
\label{eq:C-centered}
\end{equation}
Centering is a convenience, not a requirement for the $q^2$ power law.  For a noncentered scaled kernel with
\begin{equation}
{\boldsymbol\mu}(\hat{\bf q})=
\int d^2x\,{\bf x}\,k_{\hat{\bf q}}({\bf x}),
\end{equation}
the same conclusion holds with
\begin{equation}
C(\hat{\bf q})
=
|\hat{\bf q}+{\boldsymbol\mu}(\hat{\bf q})|^2
+
\operatorname{tr}\Sigma(\hat{\bf q}),
\label{eq:C-noncentered}
\end{equation}
where
\begin{equation}
\Sigma_{ij}(\hat{\bf q})
=
\int d^2x\,
[x_i-\mu_i(\hat{\bf q})][x_j-\mu_j(\hat{\bf q})]
k_{\hat{\bf q}}({\bf x}).
\end{equation}

Equation~\eqref{eq:theorem} is the central momentum-resolution result.  A nonnegative kernel with width $O(q)$ renormalizes the prefactor of the conserved density response while leaving its leading $q^2$ power intact.  The positivity assumption means that Eq.~\eqref{eq:measured} represents incoherent averaging of spectral weight.  Coherent interference, matrix-element zeros, Umklapp-shifted components, and geometry-dependent electrodynamic inversions are distinct mechanisms and require separate modeling.

The proof is given in Appendix~\ref{app:proof}.  The proof uses only positivity, normalization, the scaled form of the kernel, the second-moment/tight-tail conditions, the uniform $q^2$ asymptotics, and the technical ultraviolet domination condition.  No analyticity beyond that already contained in the fixed-frequency long-wavelength limit is required.

\section{Gaussian corollary and fixed-width contrast}
\label{sec:gaussian}

For an isotropic Gaussian resolution function with width proportional to $q$,
\begin{equation}
k_{\hat{\bf q}}({\bf x})=\frac{1}{2\pi\sigma^2}
\exp\!\left(-\frac{|{\bf x}|^2}{2\sigma^2}\right),
\label{eq:gaussian}
\end{equation}
the centering and tight-tail hypotheses are automatic, and
\begin{equation}
M_2=2\sigma^2,
\qquad
C=1+2\sigma^2.
\end{equation}
Therefore
\begin{equation}
\chi''_{\rm meas}(q\hat{\bf q},\omega)
=q^2(1+2\sigma^2)S(\omega)+o(q^2).
\label{eq:gaussian-corollary}
\end{equation}
Thus any Gaussian broadening law whose width is asymptotically proportional to $q$ preserves the Ward-identity $q^2$ suppression.  If an in-plane Gaussian broadening law is extrapolated to small momentum as $\Delta q_h(q)=\sigma q$, it belongs to the kernel class above.

\begin{figure}[ht]
\centering
\includegraphics[width=0.95\textwidth]{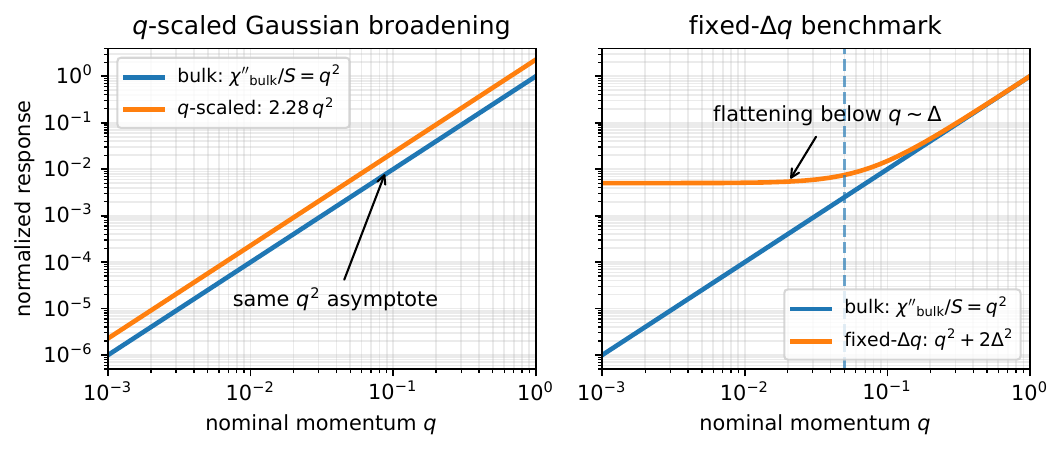}
\caption{Illustration of the momentum-resolution theorem.  Left: for a toy bulk law $\chi''_{\rm bulk}/S=q^2$, $q$-scaled Gaussian broadening with per-component width $\sigma q$ preserves the same $q^2$ asymptotic behavior and only rescales the prefactor.  The plotted value is $\sigma=0.8$, giving $1+2\sigma^2=2.28$.  Right: an illustrative fixed-$\Delta q$ Gaussian benchmark with per-component width $\Delta$ instead produces low-$q$ flattening below $q\sim\Delta$.  The theorem applies to the left-hand class of models; the fixed-$\Delta q$ comparison is heuristic.}
\label{fig:theorem}
\end{figure}

Figure~\ref{fig:theorem} provides a physical summary of the theorem.  The left panel shows the theorem's main message: when the sampled momenta remain of order the nominal momentum, averaging changes only the coefficient multiplying the conserved $q^2$ law.  The right panel shows the contrasting intuition for an absolute resolution scale: if the apparatus or sample supplies a fixed momentum width, the lowest nominal momenta sample a finite cloud of momenta and can look flat until $q$ exceeds that width.

The two panels therefore represent physically different small-$q$ limits.  In the left panel the momentum-resolution width collapses with the nominal momentum, so the measurement approaches the same conserved mode as $q\to0$.  In the right panel the width remains finite, so the nominal $q\to0$ measurement is not a measurement of the $Q\to0$ response alone.  This distinction is why fixed-width broadening is a crossover mechanism rather than a counterexample to the theorem.

A useful comparison is a fixed absolute broadening scale, $\Delta q$ independent of nominal momentum.  Such a fixed-width kernel is distinct from the scaled form in Eq.~\eqref{eq:qscaled-kernel}.  It can mask the small-$q$ regime over a finite window because, for nominal $q\ll\Delta q$, the measurement samples momenta of order $\Delta q$ rather than momenta of order $q$.  In the simplest toy model $\chi''_{\rm bulk}({\bf Q},\omega)/S(\omega)=|{\bf Q}|^2$, Gaussian averaging with fixed width $\Delta$ gives
\begin{equation}
\frac{\chi''_{\rm meas}(q\hat{\bf q},\omega)}{S(\omega)}=q^2+2\Delta^2,
\label{eq:fixed-width-toy}
\end{equation}
which is flat below $q\sim\Delta$.  This fixed-width benchmark is useful as intuition, and it is distinct from the theorem because the kernel is not self-similar under $q\to0$.  It introduces an independent absolute momentum scale and therefore predicts a crossover tied to that scale.

\section{EELS observables and electrodynamic conversion}
\label{sec:eels}

We formulate the result in terms of the bulk proper charge-density response.  EELS experiments detect energy loss by an external electron, and the measured intensity depends on Coulomb propagation, dielectric screening, form factors, surface boundary conditions, and the experimental geometry.  In a schematic bulk description one often writes a loss function in terms of a longitudinal dielectric function, while in reflection geometries one must instead use a surface response function~\cite{Egerton2011,Vig2017,Chen2024,deVries2025}.  These distinctions matter for powers of $q$.

Table~\ref{tab:objects} is the response-function map used in the rest of the paper.  The first row is the proper bulk charge polarization constrained directly by the Ward identity.  The remaining rows are the conductivity, screened, bulk-loss, and surface-loss objects through which optical and EELS measurements are commonly discussed.  Keeping these entries separate is the simplest way to see why a nearly momentum-independent loss feature need not by itself contradict the $q^2$ scaling of the proper density response.

\begin{table}[t]
\caption{Long-wavelength objects relevant to optics and EELS.  The theorem is applied to the proper bulk charge-density polarization, or to a positive $q$-scaled convolution of it; raw loss intensity requires the corresponding electrodynamic conversion.}
\small
\begin{tabular}{@{}>{\raggedright\arraybackslash}p{0.25\textwidth}>{\raggedright\arraybackslash}p{0.31\textwidth}>{\raggedright\arraybackslash}p{0.36\textwidth}@{}}
\toprule
Object & Symbol & Small-$q$ status \\
\midrule
Proper bulk charge polarization & $\Pi_{\rho\rho}$ or $\chi_{\rho\rho}^{\rm prop}$ & $\propto q^2$ if $\sigma_L$ has a finite optical limit \\
Longitudinal matter conductivity & $\sigma_L$ & finite optical limit by assumption \\
Screened bulk response & $\chi_{\rm ext}=\Pi/(1-V\Pi)$, schematically & Coulomb-kernel dependent \\
Bulk loss function & $-\operatorname{Im}\epsilon_L^{-1}$ & set by density response and Coulomb kernel \\
Surface/reflection loss response & $g({\bf q},\omega)$ or geometry-specific kernel & boundary-condition and geometry dependent \\
\bottomrule
\end{tabular}
\label{tab:objects}
\end{table}

The table also indicates where the theorem enters an experimental analysis.  If the electrodynamic conversion produces the first-row object, or a positive $q$-scaled convolution of that object, the $q^2$ theorem applies directly.  If the observed quantity is instead a screened loss function or a surface response, the extra Coulomb and boundary-condition factors must be retained before any comparison of powers of $q$ is made.

Consequently, comparison with experiment is most naturally organized in two steps.  First, an electrodynamic model maps the measured loss signal to a bulk variable.  Second, when that variable is the proper conserved density response, or a positive $q$-scaled convolution of it, the theorem fixes the long-wavelength scaling.  Surface modes, screened loss functions with explicit Coulomb factors, fixed out-of-plane momentum sampling, and responses involving an absolute momentum scale are handled at the first step rather than by the convolution theorem itself.

This distinction also clarifies the role of plasmons.  A low-momentum plasmonic regime can be entirely consistent with optical spectroscopy because Coulomb interactions reorganize the density response into a collective longitudinal electromagnetic excitation.  The proper density spectral weight entering the longitudinal conductivity remains $q^2$ suppressed at fixed nonzero frequency as long as the regular $q\to0$ longitudinal conductivity is finite.  The finite-$q$ incoherent continuum then either crosses over to this conserved regime or is described by additional surface, screening, or inhomogeneous ingredients in the experimental response.

\section{Crossover scale \texorpdfstring{$q_\ast(\omega,T)$}{q*(omega,T)}}
\label{sec:crossover}

The matching analysis identifies a crossover when a nearly momentum-independent continuum is observed over a finite momentum range but the local optical limit is metallic.  Before assuming any power laws, the crossover scale is fixed by matching the nearly local finite-$q$ density spectral weight to the conserved long-wavelength form:
\begin{equation}
\chi''_{\rm loc}(\omega,T)
\sim
q_\ast^2(\omega,T)S(\omega,T),
\end{equation}
where
\begin{equation}
S(\omega,T)=
\frac{\operatorname{Re}\sigma_{\rm opt}(\omega,T)}{\omega}.
\end{equation}
Thus
\begin{equation}
q_\ast^2(\omega,T)
\sim
\frac{\chi''_{\rm loc}(\omega,T)}{S(\omega,T)}
=
\frac{\omega\,\chi''_{\rm loc}(\omega,T)}
{\operatorname{Re}\sigma_{\rm opt}(\omega,T)}.
\label{eq:qstar-finiteT}
\end{equation}
This relation is a phenomenological matching estimate that follows after the long-wavelength asymptotic and finite-$q$ regime have both been identified.

As an illustrative zero-temperature or fixed-temperature frequency-window specialization, suppose that the preasymptotic finite-$q$ continuum has the form
\begin{equation}
\chi''_{\rm loc}(q,\omega)\sim C_>\omega^{-\alpha},
\label{eq:local-regime}
\end{equation}
where the leading term is approximately independent of momentum.  Suppose also that the true conserved low-$q$ response has
\begin{equation}
\chi''_{\rm cons}(q,\omega)
\sim C_< q^2\omega^{-1-\beta},
\label{eq:conserved-regime}
\end{equation}
corresponding to $\operatorname{Re}\sigma_{\rm opt}(\omega)\sim \omega^{-\beta}$.  Matching the two estimates at $q=q_\ast(\omega)$ gives
\begin{equation}
q_\ast(\omega)
\sim
\left(\frac{C_>}{C_<}\right)^{1/2}
\omega^{(1+\beta-\alpha)/2}.
\label{eq:qstar-general}
\end{equation}
For the often discussed combination of a frequency-flat finite-$q$ continuum, $\alpha=0$, and an optical conductivity scaling $\operatorname{Re}\sigma(\omega)\sim\omega^{-2/3}$, representative of power-law optical analyses in cuprates~\cite{vanDerMarel2003,vanDerMarel2006}, one obtains
\begin{equation}
q_\ast(\omega)\propto\omega^{5/6}.
\label{eq:qstar-fivesixths}
\end{equation}

A finite-temperature scaling form can be treated similarly.  If the finite-$q$ continuum has
\begin{equation}
\chi''_{\rm loc}(\omega,T)=T^{-\nu}f(\omega/T)
\end{equation}
and the optical conductivity has
\begin{equation}
\operatorname{Re}\sigma_{\rm opt}(\omega,T)=T^{-\beta_\sigma}g(\omega/T),
\end{equation}
then Eq.~\eqref{eq:qstar-finiteT} gives
\begin{equation}
q_\ast^2(\omega,T)
\sim
\omega T^{\beta_\sigma-\nu}
\frac{f(\omega/T)}{g(\omega/T)}.
\label{eq:qstar-scaling}
\end{equation}
At fixed $\omega/T$, this implies $q_\ast\propto T^{(1+\beta_\sigma-\nu)/2}$ up to the scaling functions.  This form is useful when discussing scale-invariant strange-metal fits such as those reported in Ref.~\cite{Guo2024}.

The scaling estimate should be used with the same response-function bookkeeping as the theorem.  The powers $\alpha$, $\beta$, $\nu$, and $\beta_\sigma$ need only describe an intermediate frequency or temperature window; the theorem itself is pointwise in $\omega$ and assumes only the regular long-wavelength conductivity limit.  Weak residual momentum dependence of the finite-$q$ continuum gives corresponding corrections to Eq.~\eqref{eq:qstar-finiteT}.  For loss-function data, the matching is performed after conversion to the corresponding bulk density variable.

\subsection{Candidate physical origins of the crossover}
\label{sec:qstar-mechanisms}

When a proper finite-$q$ continuum connects to a regular optical metal, the theorem requires a crossover but does not compute the microscopic value of $q_\ast$.  Several intrinsic bulk mechanisms can contribute to such a scale.

First, any nonlocal transport length produces a natural momentum boundary.  In a diffusive metal this length is the diffusion length $L_\omega\sim(D/\omega)^{1/2}$, giving a crossover when $qL_\omega\sim1$.  In a more incoherent metal the analogous scale may be set by a relaxation length or by a characteristic velocity times a relaxation time.  This mechanism does not require quasiparticles; it only requires that the local optical response cease to be adequate once the probe varies appreciably over the relevant relaxation length.

Second, layered Coulomb electrodynamics can separate the low-momentum plasmonic regime from a more microscopic finite-momentum regime.  Interlayer Coulomb coupling, the distinction between in-phase and out-of-phase charge oscillations, and entry into a particle-hole continuum can change the plasmon dispersion and damping without altering the Ward constraint on the proper long-wavelength density response~\cite{Nucker1991,Grigoryan1999,deVries2025,Schultz2025}.  In this case $q_\ast$ is tied to an intrinsic electrodynamic or band-structure scale, not to experimental resolution.

Third, strange-metal fluctuations may introduce a correlation length or dynamic scaling scale.  If the finite-$q$ continuum is governed by a critical or locally critical sector, the crossover can occur when $q\xi(\omega,T)$ becomes of order one, or equivalently when the spatial part of the scaling function becomes important.  Equation~\eqref{eq:qstar-scaling} is a phenomenological version of this idea.

Finally, lattice-scale physics, Umklapp processes, interband matrix elements, short-range charge correlations, and pseudogap or charge-order fluctuations can all change the finite-momentum spectrum once $q$ is no longer asymptotically small.  These mechanisms are intrinsic if they enter the bulk proper response itself.  They are distinct from extrinsic fixed-width resolution or surface-response effects, although both intrinsic and extrinsic mechanisms can produce finite-window flattening in measured spectra.

\begin{figure}[ht]
\centering
\includegraphics[width=0.62\textwidth]{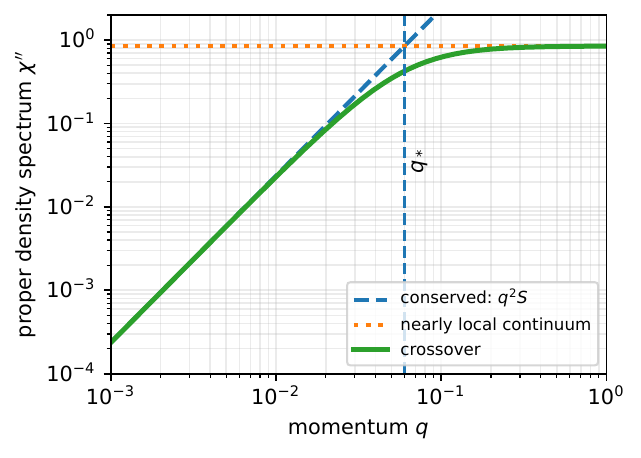}
\caption{Schematic crossover implied by charge conservation.  At fixed nonzero frequency and temperature, a homogeneous conserved metallic response must satisfy $\chi''_{\rm bulk}\sim q^2S(\omega,T)$ as $q\to0$.  A nearly local finite-$q$ continuum can exist over an intermediate momentum window, but if it connects to the optical metal it must cross over at $q_\ast(\omega,T)$.}
\label{fig:crossover}
\end{figure}

Figure~\ref{fig:crossover} summarizes the role of these mechanisms.  The Ward identity fixes the left side of the figure, while microscopic physics and electrodynamic conversion determine where the curve leaves the conserved $q^2$ regime and enters the finite-$q$ continuum.

\section{Implications for Bi-2212 and bad metals}
\label{sec:experiment}

The combined experimental picture is best viewed as a crossover problem rather than a yes-or-no question about plasmons.  Early transmission EELS measurements on cuprates reported dispersive long-wavelength collective excitations and were analyzed in terms of conventional plasmon physics~\cite{Nucker1991,Grigoryan1999}.  More recent M-EELS work reported a broad continuum in the strange-metal regime~\cite{Mitrano2018,Husain2019}, leading to the subsequent Comment/Reply exchange on momentum calibration, resolution, and the interpretation of the large-$q$ spectra~\cite{Fink2021,HusainReply2021}.  The recent review by Abbamonte and Fink presents this as a broader issue in EELS studies of correlated materials~\cite{AbbamonteFink2025}.

Recent Bi-2212 measurements sharpen the same point.  Reflection M-EELS reports a plasmon feature quantitatively consistent with infrared optics for $q<0.04$ reciprocal lattice units and an incoherent, marginal-Fermi-liquid-like continuum for $q>0.04$ reciprocal lattice units~\cite{Chen2024}.  Transmission EELS reports measurements with $\Delta E\approx30$ meV and $\Delta q\approx0.01\,\text{\AA}^{-1}$, a highly damped plasmonic excitation for $q<0.15\,\text{\AA}^{-1}$, and an incoherent continuum for $q>0.15\,\text{\AA}^{-1}$~\cite{deVries2026}.  Low-energy measurements also report nearly separable, scale-invariant behavior with weak momentum dependence~\cite{Guo2024}; if interpreted as a proper density response, such behavior must still be related to the lowest-momentum electrodynamic regime.  These reported momentum scales are empirical candidates for the crossover scale discussed above, once the measured loss signal has been converted to the appropriate bulk variable.

The same framework is compatible with observations of well-defined plasmons in other bad or correlated metals.  For example, transmission-EELS measurements on Sr$_2$RuO$_4$ report well-defined optical and acoustic plasmons in a material that crosses into a bad-metal regime~\cite{Schultz2025}.  Such observations do not conflict with the Ward constraint; a plasmon is a Coulomb-dressed collective mode built from a proper density response whose small-$q$ spectral weight remains constrained by charge conservation.  The framework therefore accommodates both coherent low-$q$ plasmons and incoherent finite-$q$ continua, provided the response function and the order of limits are specified.

The present result clarifies how to interpret the separation between low-$q$ and finite-$q$ regimes.  Rather than requiring pointwise equality between optical spectroscopy and EELS, it constrains any homogeneous conserved bulk longitudinal response that connects the local optical limit to the finite-$q$ continuum.  If the corresponding long-wavelength longitudinal conductivity is finite, the bulk proper density spectral weight vanishes as $q^2$ at fixed nonzero frequency.  A nearly local finite-$q$ continuum is then naturally interpreted as a preasymptotic momentum regime, or as part of a measured response that also contains an absolute momentum scale, surface or geometry-specific loss effects, fixed out-of-plane momentum sampling, or other ingredients beyond the homogeneous conserved-bulk convolution model.

The most useful experimental target is therefore the crossover surface $q_\ast(\omega,T)$.  At each fixed nonzero $\omega$ and $T$, after converting the measured loss signal to the proper bulk density variable, the lowest accessible momenta should show a $q^2$ density spectral weight if the optical conductivity is finite.  The momentum where this behavior gives way to the nearly local continuum defines $q_\ast(\omega,T)$.  Measuring this surface is more diagnostic than asking whether spectra are momentum independent over a finite window.  It would directly test whether the observed finite-$q$ continuum can be matched to the optical conductivity by Eq.~\eqref{eq:qstar-finiteT}, or whether additional electrodynamic, finite-resolution, or microscopic ingredients beyond the simple homogeneous-bulk matching are required.

\section{Discussion and conclusion}
\label{sec:discussion}

Charge conservation imposes a strong long-wavelength constraint on longitudinal density fluctuations.  At fixed nonzero frequency, a homogeneous metal with a finite long-wavelength longitudinal matter conductivity has a proper charge-density spectral function that vanishes as $q^2$.  This statement is independent of microscopic quasiparticles, Boltzmann transport, or any particular theory of the strange metal.

The momentum-resolution theorem extends this constraint to experimentally motivated averaging models.  A broad class of nonnegative momentum-resolution mechanisms with self-similar $q$-scaled profiles and controlled second moments preserves the $q^2$ asymptotics.  Such mechanisms alter amplitudes and shapes in a finite momentum window while leaving the leading long-wavelength power law unchanged.  Gaussian $q$-scaled broadening is a simple corollary.

Fixed absolute broadening gives a different phenomenology.  It can obscure the asymptotic regime by sampling momenta of order the fixed width even when the nominal momentum is smaller.  This introduces an independent momentum scale and points to flattening below a scale set by the broadening, rather than to an asymptotic local proper density response.

For strange-metal Bi-2212 and related bad metals, the conclusion is conservative but sharp.  Coherent low-momentum plasmons, overdamped plasmons, and nearly local finite-$q$ continua are all possible finite-window phenomena once Coulomb electrodynamics and microscopic damping are included.  What charge conservation fixes is the strict long-wavelength limit of the proper homogeneous density response.  If that response connects to a finite optical conductivity, it must cross over to the $q^2$ optical-limit form at sufficiently small momentum.  The necessary object to determine experimentally and theoretically is the crossover surface connecting the optical metal to the finite-$q$ continuum.

\section*{Acknowledgments}
The author thanks Ryumi S.\ for inspiration, Damian G.\ and Chris M.\ for useful discussions, and George Sterman for his lasting intellectual influence.  The views expressed are the author's and do not necessarily represent the U.S. Department of Homeland Security or the United States government.

\appendix

\section{Ward-identity conventions}
\label{app:ward}

This appendix records the sign and source conventions used in Sec.~\ref{sec:ward}.  We use the physical charge density $\rho$ and charge current ${\bf J}$, with continuity equation
\begin{equation}
\partial_t\rho+\nabla\cdot{\bf J}=0.
\end{equation}
The scalar source $\Phi$ is defined operationally as the total longitudinal electric potential for which
\begin{equation}
\delta\rho({\bf q},\omega)=\chi^R_{\rho\rho}({\bf q},\omega)\Phi({\bf q},\omega).
\end{equation}
Equivalently, if one writes an explicit perturbing Hamiltonian, its sign must be chosen consistently with this operational definition.  Nothing in the theorem depends on the name assigned to the scalar source; the invariant content is the relation between dissipative density response and matter conductivity.

With Fourier convention $e^{i({\bf q}\cdot{\bf r}-\omega t)}$, the longitudinal electric field is
\begin{equation}
E_i({\bf q},\omega)=-iq_i\Phi({\bf q},\omega),
\end{equation}
and the continuity equation gives
\begin{equation}
-i\omega\delta\rho+iq_iJ_i=0,
\qquad\text{or}\qquad
\omega\delta\rho=q_iJ_i.
\end{equation}
Using $J_i=\sigma_{ij}E_j$, one obtains
\begin{equation}
\chi^R_{\rho\rho}\Phi
=\frac{q_i}{\omega}\sigma_{ij}E_j
=-\frac{i q_iq_j}{\omega}\sigma_{ij}\Phi
=\frac{q_iq_j}{i\omega}\sigma_{ij}\Phi.
\end{equation}
For an isotropic response, this is Eq.~\eqref{eq:ward}; for an anisotropic response it gives Eq.~\eqref{eq:tensor-exact}.  Writing $\sigma_L=\sigma_1+i\sigma_2$ gives
\begin{equation}
\operatorname{Im}\chi^R_{\rho\rho}=-\frac{q^2}{\omega}\sigma_1,
\end{equation}
so for $\omega>0$ the positive spectral function $\chi''_{\rm bulk}=-\operatorname{Im}\chi^R_{\rho\rho}$ obeys Eq.~\eqref{eq:exact}.  A different overall sign convention for the scalar source would reverse the sign of $\chi^R_{\rho\rho}$ and of the corresponding definition of $\chi''$; Eq.~\eqref{eq:exact}, which relates the positive dissipative density spectrum to $\operatorname{Re}\sigma_L$, is the convention used throughout the paper.

If the number density is used instead of the charge density, $\rho=-en$.  A scalar electric potential $\Phi$ coupling to $\rho$ corresponds to a number-density source $\phi_n=-e\Phi$.  Therefore
\begin{equation}
\chi^R_{\rho\rho}=e^2\chi^R_{nn},
\qquad
\chi''_{\rho\rho}=e^2\chi''_{nn},
\end{equation}
provided the sources are converted consistently.  This is why factors of $e$ do not affect the small-$q$ Ward power.

\section{Proper, screened, and loss-function responses}
\label{app:screened}

The Ward identity constrains the proper response related to the matter conductivity.  In a charged system, the density induced by an externally imposed potential is obtained only after solving the electrostatic or electrodynamic self-consistency problem.  The object constrained in this paper is the irreducible polarization $\Pi_{\rho\rho}$, denoted $\chi_{\rho\rho}$ in the main text, not the fully screened density response to a distant external probe.  Equivalently, $\chi_{\rho\rho}$ is the polarization entering $\epsilon_L=1-V_q\chi_{\rho\rho}$ before the final Coulomb resummation, rather than the external screened response after that resummation.  In a schematic translationally invariant bulk treatment one may write
\begin{equation}
\chi_{\rm ext}(q,\omega)=
\frac{\chi_{\rho\rho}(q,\omega)}{1-V_q\chi_{\rho\rho}(q,\omega)},
\label{eq:screened-schematic}
\end{equation}
up to sign conventions for the charge and potential, with a corresponding dielectric function $\epsilon_L(q,\omega)=1-V_q\chi_{\rho\rho}(q,\omega)$.  A measured bulk loss function is then related to $-\operatorname{Im}\epsilon_L^{-1}$ rather than directly to $\chi''_{\rho\rho}$.  In reflection EELS the relation is further modified by surfaces and layered electrodynamics.

Equation~\eqref{eq:screened-schematic} is included to clarify the response-function bookkeeping.  The momentum-resolution theorem is independent of the detailed Coulomb kernel or surface response function.  When the object being smeared is the homogeneous conserved bulk proper density response whose longitudinal conductivity has the regular local limit measured optically, its leading fixed-frequency small-$q$ behavior is $q^2$ and remains $q^2$ under nonnegative $q$-scaled momentum averaging.  Comparison with EELS data therefore proceeds by first identifying the electrodynamic conversion between the measured loss signal and this bulk response.

\section{Proof of the momentum-resolution theorem}
\label{app:proof}

We prove Eq.~\eqref{eq:theorem}.  Fix a compact frequency window $I\subset(0,\infty)$ and assume Eqs.~\eqref{eq:remainder}--\eqref{eq:growth}.  First note that $S(\omega)$ is bounded on $I$.  Choose any sufficiently small nonzero $Q_0$ in the regime where Eq.~\eqref{eq:remainder} applies.  Then
\begin{equation}
Q_0^2|S(\omega)|
\le |\chi''_{\rm bulk}(Q_0\hat{\bf e},\omega)|+|r(Q_0\hat{\bf e},\omega)|,
\end{equation}
and the right-hand side is uniformly bounded on $I$ by the growth bound and the uniform remainder bound.  We write
\begin{equation}
S_I=\sup_{\omega\in I}|S(\omega)|<\infty.
\end{equation}

Changing variables in Eq.~\eqref{eq:measured} according to
\begin{equation}
{\bf Q}=q(\hat{\bf q}+{\bf x})
\end{equation}
gives
\begin{equation}
\chi''_{\rm meas}(q\hat{\bf q},\omega)
=
\int_{\mathbb{R}^2}d^2x\,k_{\hat{\bf q}}({\bf x})\,
\chi''_{\rm bulk}(q(\hat{\bf q}+{\bf x}),\omega).
\label{eq:app-change-vars}
\end{equation}
Subtract the proposed leading term and divide by $q^2$:
\begin{align}
D(q,\hat{\bf q},\omega)
&=\frac{\chi''_{\rm meas}(q\hat{\bf q},\omega)}{q^2}
-S(\omega)\int d^2x\,|\hat{\bf q}+{\bf x}|^2k_{\hat{\bf q}}({\bf x})
\nonumber\\
&=\int d^2x\,k_{\hat{\bf q}}({\bf x})
\frac{r(q(\hat{\bf q}+{\bf x}),\omega)}{q^2}.
\label{eq:D}
\end{align}
It remains to show that
\begin{equation}
\sup_{\hat{\bf q}\in S^1,\,\omega\in I}|D(q,\hat{\bf q},\omega)|\to0
\qquad (q\to0).
\label{eq:Dgoal}
\end{equation}

Let $\epsilon>0$.  By Eq.~\eqref{eq:remainder}, choose $\eta>0$ such that
\begin{equation}
\sup_{\omega\in I}\frac{|r({\bf Q},\omega)|}{|{\bf Q}|^2}<\epsilon
\qquad \text{whenever } |{\bf Q}|<\eta.
\label{eq:eta-choice}
\end{equation}
For each $q$ and $\hat{\bf q}$ define
\begin{equation}
A_q(\hat{\bf q})=\bigl\{{\bf x}:|\hat{\bf q}+{\bf x}|\le\eta/q\bigr\},
\qquad
B_q(\hat{\bf q})=\mathbb{R}^2\setminus A_q(\hat{\bf q}).
\end{equation}
Write $D=D_A+D_B$ for the corresponding contributions.

On $A_q$, Eq.~\eqref{eq:eta-choice} gives
\begin{equation}
|r(q(\hat{\bf q}+{\bf x}),\omega)|
\le \epsilon q^2|\hat{\bf q}+{\bf x}|^2.
\end{equation}
Hence
\begin{align}
|D_A(q,\hat{\bf q},\omega)|
&\le \epsilon\int d^2x\,k_{\hat{\bf q}}({\bf x})|\hat{\bf q}+{\bf x}|^2
\nonumber\\
&\le 2\epsilon\int d^2x\,k_{\hat{\bf q}}({\bf x})(1+|{\bf x}|^2).
\end{align}
The uniform second-moment bound therefore gives
\begin{equation}
\sup_{\hat{\bf q},\omega}|D_A(q,\hat{\bf q},\omega)|\le C_1\epsilon
\label{eq:DA-bound}
\end{equation}
for a finite constant $C_1$ independent of $q$.

On $B_q$ one has $|\hat{\bf q}+{\bf x}|>\eta/q$.  If $q<\eta/2$, then
\begin{equation}
|{\bf x}|>\frac{\eta}{q}-1>\frac{\eta}{2q}.
\label{eq:tail-set}
\end{equation}
Using
\begin{equation}
r(q(\hat{\bf q}+{\bf x}),\omega)=
\chi''_{\rm bulk}(q(\hat{\bf q}+{\bf x}),\omega)
-q^2|\hat{\bf q}+{\bf x}|^2S(\omega)
\end{equation}
and Eq.~\eqref{eq:growth},
\begin{align}
|D_B(q,\hat{\bf q},\omega)|
&\le
\frac{A_I}{q^2}\int_{B_q}d^2x\,k_{\hat{\bf q}}({\bf x})
\nonumber\\
&\quad +(A_I+S_I)
\int_{B_q}d^2x\,k_{\hat{\bf q}}({\bf x})|\hat{\bf q}+{\bf x}|^2.
\label{eq:DB-start}
\end{align}
For the first term, Eq.~\eqref{eq:tail-set} gives
\begin{align}
\frac{1}{q^2}\int_{B_q}d^2x\,k_{\hat{\bf q}}({\bf x})
&\le
\frac{1}{q^2}\int_{|{\bf x}|>\eta/(2q)}d^2x\,k_{\hat{\bf q}}({\bf x})
\nonumber\\
&\le
\frac{4}{\eta^2}\int_{|{\bf x}|>\eta/(2q)}d^2x\,|{\bf x}|^2 k_{\hat{\bf q}}({\bf x}).
\label{eq:first-tail}
\end{align}
The right-hand side tends to zero uniformly in $\hat{\bf q}$ by the tight-tail hypothesis.  For the second term, on $B_q$ and for sufficiently small $q$ one has $|{\bf x}|>1$, and therefore
\begin{equation}
|\hat{\bf q}+{\bf x}|^2\le2(1+|{\bf x}|^2)\le4|{\bf x}|^2.
\end{equation}
Thus
\begin{equation}
\int_{B_q}d^2x\,k_{\hat{\bf q}}({\bf x})|\hat{\bf q}+{\bf x}|^2
\le
4\int_{|{\bf x}|>\eta/(2q)}d^2x\,|{\bf x}|^2 k_{\hat{\bf q}}({\bf x}),
\label{eq:second-tail}
\end{equation}
which also vanishes uniformly.  Combining Eqs.~\eqref{eq:DB-start}--\eqref{eq:second-tail},
\begin{equation}
\sup_{\hat{\bf q},\omega}|D_B(q,\hat{\bf q},\omega)|\to0
\qquad (q\to0).
\label{eq:DB-vanish}
\end{equation}
Since $\epsilon$ in Eq.~\eqref{eq:DA-bound} is arbitrary, Eqs.~\eqref{eq:DA-bound} and \eqref{eq:DB-vanish} prove Eq.~\eqref{eq:Dgoal}, and hence Eq.~\eqref{eq:theorem}.

\section{Gaussian kernels}
\label{app:gaussian}

For the isotropic Gaussian in Eq.~\eqref{eq:gaussian},
\begin{equation}
\int d^2x\,{\bf x}\,k_{\hat{\bf q}}({\bf x})=0,
\qquad
\int d^2x\,|{\bf x}|^2k_{\hat{\bf q}}({\bf x})=2\sigma^2.
\end{equation}
The second-moment tail is
\begin{equation}
\int_{|{\bf x}|>R}d^2x\,|{\bf x}|^2k_{\hat{\bf q}}({\bf x})
=
\left(R^2+2\sigma^2\right)e^{-R^2/(2\sigma^2)},
\end{equation}
which tends to zero exponentially as $R\to\infty$.  The theorem therefore applies and gives Eq.~\eqref{eq:gaussian-corollary}.  The prefactor is obtained directly:
\begin{align}
C
&=\int d^2x\,|\hat{\bf q}+{\bf x}|^2 k_{\hat{\bf q}}({\bf x})
\nonumber\\
&=1+2\hat{\bf q}\cdot\int d^2x\,{\bf x}k_{\hat{\bf q}}({\bf x})+
\int d^2x\,|{\bf x}|^2 k_{\hat{\bf q}}({\bf x})
\nonumber\\
&=1+2\sigma^2.
\end{align}

\section{Fixed absolute width as a benchmark}
\label{app:fixedwidth}

For comparison, consider a normalized isotropic Gaussian with fixed width $\Delta$ in absolute momentum units,
\begin{equation}
K_{\Delta,q\hat{\bf q}}({\bf Q})=
\frac{1}{2\pi\Delta^2}
\exp\!\left[-\frac{|{\bf Q}-q\hat{\bf q}|^2}{2\Delta^2}\right].
\end{equation}
If the toy bulk response is exactly $\chi''_{\rm bulk}({\bf Q},\omega)=S(\omega)|{\bf Q}|^2$, then
\begin{align}
\chi''_{\rm meas}(q\hat{\bf q},\omega)
&=S(\omega)\int d^2Q\,K_{\Delta,q\hat{\bf q}}({\bf Q})|{\bf Q}|^2
\nonumber\\
&=S(\omega)(q^2+2\Delta^2),
\end{align}
which is Eq.~\eqref{eq:fixed-width-toy}.  This calculation illustrates why a fixed-width mechanism can flatten a nominal small-$q$ measurement.  It also shows why such a mechanism is conceptually distinct from the $q$-scaled theorem: the underlying conserved $q^2$ asymptotics remains, while the measurement samples momenta of order $\Delta$ even when the nominal momentum is smaller.

\end{document}